# Evolving a Model for Software Process Context: An Exploratory Study

Diana Kirk[1] and Stephen G. MacDonell[2]
[1]*Technology Academy, EDENZ Colleges, 85 Airedale Street, Auckland 1010, New Zealand*
[2]*Software Engineering Laboratory (SERL), AUT University, Private Bag 92006, Auckland 1142, New Zealand*

**Abstract**

*In the domain of software engineering, our efforts as researchers to advise industry on which software practices might be applied most effectively are limited by our lack of evidence based information about the relationships between context and practice efficacy. In order to accumulate such evidence, a model for context is required. We are in the exploratory stage of evolving a model for context for situated software practices. In this paper, we overview the evolution of our proposed model. Our analysis has exposed a lack of clarity in the meanings of terms reported in the literature. Our base model dimensions are People, Place, Product and Process. Our contributions are a deepening of our understanding of how to scope contextual factors when considering software initiatives and the proposal of an initial theoretical construct for context. Study limitations relate to a possible subjectivity in the analysis and a restricted evaluation base. In the next stage in the research, we will collaborate with academics and practitioners to formally refine the model.*

## 1. INTRODUCTION

In the domain of software engineering (SE), evidence suggests that practitioners adapt development methodologies to suit specific contexts (Avison and Pries-Heje, 2008; MacCormack et al., 2012; Petersen and Wohlin, 2009a; Turner et al., 2010). Moreover, research indicates that most organisations adapt practices drawn from several approaches, often at the level of the individual project. As an example, as agile approaches have become more established, limitations have been exposed, leading to either contextualisation (Campanelli and Parreiras, 2015) or amalgamation with other paradigms, for example, the 'lean' paradigm (Wang et al., 2012). In addition to the issue of tailoring, the emergence of new paradigms, for example, 'continuous delivery' (Dingsøyr and Lassenius, 2016; Stuckenberg and Heinzl, 2010), has created a need for extended and modified approaches.

This raises important questions about when and how adaptation is appropriate. Lengnick-Hall and Griffith point out that, if the intention is to achieve "a specific, designated outcome", as is the case for most software practices, the knowledge (practice) must be applied as-is. Applying the knowledge in an intuitive or experimental way introduces a lack of fit be- tween type of knowledge and how it is applied, and this inevitably leads to reduced effectiveness, at best (Lengnick-Hall and Griffith, 2011). From this perspective, the now-popular approach towards tailoring within industry might be viewed as a 'hidden' issue.

To avoid the potential problems inherent in ad- hoc adaptation, it is thus crucial that the tailoring is understood within the context in which it will be applied. A theoretical model for context is required. Indeed, lack of a defined construct for context has in the past created problems for researchers and practitioners. In the first instance, researchers who carry out formal experiments are unable to confidently interpret the scope of applicability of their results because the role of contextual factors is insufficiently understood (Basili et al., 1999; Carver et al., 2004; Runeson et al., 2014; Sjøberg et al., 2005). Second, there is inherent uncertainty in the use of available data repositories for investigation and estimation (for example, estimation of project effort) because the environment associated with the data is at best only partially stated (Bosu and MacDonell, 2013).

Our goal is to support researchers in the accumulation of context-related evidence to be used as a basis for evidence-based-software engineering (EBSE). Our vision is that, as researchers understand the kinds of information that need to be captured as 'context', growing evidence repositories will yield 'practice families' i.e. sets of similar practices that are indicated (and contra-indicated) for a specific value along one dimension. 'Best practice' will then in- volve choosing from practices that are consistent with values along all dimensions i.e. either indicated or not-contra-indicated. A *lean* process is one in which choices represent an overall maximal effectiveness.

Our journey will comprise three phases, a) an exploratory phase to evolve a candidate model, b) a refinement stage where we formally refine the model, and c) an application phase where we generate and test hypotheses based on the model (Routio, 2007). In this paper, we describe our approach to, and implementation of, the exploratory phase i.e. initial model generation.

In section 2, we overview related work. In section 3, we present our research approach. In section 4 we describe the evolution of our model. In section 5, we discuss some limitations of both the study and the proposed model and in section 6, we summarise the paper and discuss future work.

## 2. RELATED WORK

There are two areas of related work for this paper. The first includes research aimed at more informal efforts to categorise context along various dimensions. The second includes attempts to provide suitable theoretical constructs for context. Space limitations enable us to present an overview only.

### 2.1 Context Models

There have been many efforts to relate SE outcomes to specific key factors. We overview a selection here. Avison and Pries-Heje aimed to support selection of a suitable methodology that is project-specific (Avison and Pries-Heje, 2008). For a given project, the authors plotted position along each of eight dimensions on a radar graph and inferred an appropriate methodology from the shape of the graph. We see two limitations. First, the abstraction is based on a specific organisation, resulting in missing contexts, for example, temporal distance. Second, it is based at the level of the *project* and so is inapplicable to, for example, a 'customer-driven' environment, where the on-going relationship between development group and customer becomes key (Dingsøyr and Lassenius, 2016; Munezero et al., 2017; Stuckenberg and Heinzl, 2010).

Clarke and O'Connor propose a reference framework for situational factors affecting software development (Clarke and O'Connor, 2012). The framework includes eight classifications: *Personnel*, *Requirements*, *Application*, *Technology*, *Organisation*, *Operation*, *Management* and *Business*, further divided into 44 factors. Our critique of this approach is that the *meanings* assigned to sub-factors do not represent a consistent set with respect to practice suitability. For example, the factor 'Cohesion' includes "team members who have not worked for you", "ability to work with uncertain objectives" and "team geographically distant", each of which might indicate different kinds of practice. The framework may indeed provide a comprehensive list of factors. However, the approach remains discrete in nature and is unsuitable for classifying factors in a theoretical way, as the categories are semantically inconsistent and there are no clear rules on which to base abstraction.

Petersen and Wohlin provide a checklist for representing context for the purpose of aggregating studies in industrial settings (Petersen and Wohlin, 2009b). The facets of the structure include *Product*, *Processes*, *Practices*, *People*, *Organisation* and *Market*. The facets and context elements are presented as a given, without justification. While likely useful, there appear to be missing contexts, for example, relating to cultural mis-matches between and within teams.

### 2.2 SE Theory Building

Sjøberg et al. propose a framework that includes *Software system*, which "may be classified along many dimensions, such as size, complexity, application do- main, ..." (Sjøberg et al., 2008). The form of possible classifications for context is not discussed.

Dybå et al. observe that most empirical SE research to date has adopted a *discrete* perspective, i.e. examining "specific contextual variables" and state the need for a broader *omnibus* perspective (Dybå et al., 2012). The authors suggest that such a perspective involves a consideration of *Who*, *Where*, *What*, *When*, *How* and *Why*. The meanings of the dimensions are adapted from organisational science. We agree there is a need for a shift in focus from identifying factors to abstracting the problem space and have adopted the suggested structure as a starting point for our investigations (Kirk and MacDonell, 2014b; Kirk and MacDonell, 2014a).

An endeavour that aims to develop a general theory for software development is the SEMAT initiative (Jacobson et al., 2013). The approach proposes a set of top-level ideas that support determination of a project's health and are *Requirements*, *Software System*, *Work*, *Team*, *Way of Working*, *Opportunity* and *Stakeholders*. The approach is potentially useful if applied to a specific kind of project in which the notion of 'health' is compatible with the measures provided. However, we believe there are many kinds of project that are unsuitable for this model. In addition, there appears to be no justification for, or evaluation of, the factors as a minimal, spanning set for the space of all contexts.

Bjarnason et al. identify the various kinds of *Distance* encountered in software engineering and pro- pose a theory that supports selection of suitable practices for reducing distances (Bjarnason et al., 2015). We observe that the space of contexts is greater than that of distances and some distances may be less important when the objectives for the project are taken into account.

## 3. RESEARCH APPROACH

Routio describes three kinds of research as a) there is no model to use as a starting point (exploratory research), b) an existing model is being expanded or refined, and c) hypotheses based on an established model are being tested. Exploratory research is appropriate for a "phenomenological pursuit into deep understanding" and where there is "distrust on earlier descriptions". The researcher begins with a "preliminary notion" of the object of study. During the study the "provisional concepts ... gradually gain precision" until a suitable conceptualisation is achieved. Routio suggests that the journey may involve some "creative innovation" (Routio, 2007).

Our objective was to create an initial framework for software process context. Although there exist several proposed frameworks for context, we rejected these for two reasons. First, none emphasises the *properties* that define category membership and so categorisations are inconsistent from a *meaning* perspective (see section 2). Second, we were concerned that the result would not be sufficiently general given the fast-changing nature of software development. For example, new paradigms such as software-as-a-service (Stuckenberg and Heinzl, 2010) and continuous value delivery (Dingsøyr and Lassenius, 2016) have raised the need to rethink software process.

Rather than create a model from the literature, we believed a more conceptual approach would result in a more comprehensive model. Our goal is to evolve an initial model that will in the future be refined and then applied for hypothesis testing. We scoped our research as relating to a *software initiative* which we define as "any endeavour that involves defining, creating, delivering, maintaining or supporting software intensive products or services".

According to Creswell, the first step in any research initiative is to expose philosophical assumptions by identifying the *philosophical worldview* adopted by the researcher (Creswell, 2014). Four popular world views are *Postpositivist* (which generally concerns causation and hypothesis-testing), *Constructivist* (where the complexity of individuals' viewpoints is of interest), *Transformative* (which focuses on political effects on minority groups), and *Pragmatism* (where researchers equate truth with 'what works' and use any means available to under- stand the problem) (Creswell, 2014). For this research, we adopted a *pragmatic* worldview. The pragmatist considers theories as "the products of a consensual process ... to be judged for their utility" (Easter- brook et al., 2008). This viewpoint involves a focus on what works and supports the use of all available approaches to better understand the problem space. We applied a mix of approaches, implemented in a pragmatic way.

## 4. MODEL EVOLUTION

In table 1, we overview the activities carried out during the evolution of our proposed framework.

We based our initial structure for context on existing ideas (Dybå et al., 2012; Zachman, 2009). The structure included the dimensions *why*, *who*, *where*, *what*, *when* and *how* with meanings based on the work of Orlikowski (Orlikowski, 2002). The second step involved a small pilot where we categorised into the structure contextual factors named in three software engineering literature studies. We wanted to test that our conceptualisation represented "a starting point (e.g. a framework) that identifies aspects of a topic" (Stol and Fitzgerald, 2015). This step resulted in several findings (Kirk and MacDonell, 2014b; Kirk and MacDonell, 2014a). First, we found huge issues with terminology, a problem more recently addressed by Clarke et al., who suggest that the "proliferation of language and term usage" warrants the establishment of an ontological model for software process terminology (Clarke et al., 2016). This is a position we agree with and have explored in relation to some soft- ware process constructs (Kirk and MacDonell, 2016).

Second, we realised that named terms related to different *kinds* of context i.e. had different meanings. This was a crucial discovery — if we are to produce a model that can aid understanding of situated software practices, our starting assumption is that different kinds of contextual factor will play differerent roles and so we must establish rules for category inclusion. For example, named factors related to *organisation level strategies* (for example, 'globalise'), *project objectives* (for example, 'user acceptance'), aspects of the *process* (for example 'tool support') and *local operational context* (for example, 'developer experience'). We also understood that that our dimensional structure refers to *local operational context* i.e. we are always interested in local effect. A factor such as 'globalise' certainly may have an impact on development, but in an *indirect* way, for example, by causing teams to be set up remotely. These then become the local context for the project.

Table 1: Steps in model evolution.

| Activity | Source |
|---|---|
| Initial concept | Prior work |
| Pilot categorisation | Literature studies |
| Extend framework | Results of pilot |
| Literature categorisation | Literature studies |
| Small evaluation | Industry projects |

Table 2: Numbers of included documents.

| | Scopus | EBSCO |
|---|---|---|
| Found | 3,150 | 471 |
| Candidates | 332 | 25 |
| Included | 40 | 10 |

A third issue exposed was that some named factors are not sufficiently defined for practice tailoring. For example, 'user participation' may mean the user helped in requirements definition, is available throughout the project or carried out beta testing. 'Non-colocated team' may mean testers are in a different country or the team is split between two rooms in the same building. 'Company size' may involve all of constraints on local process choices, staff satisfaction levels and locational organisation. We introduced the terms *Secondary* for factors that are multi-dimensional and *Ambiguous* for ill-defined factors.

The key ideas that had emerged after this second step were that a) we need to be clear whether a factor relates to strategy, objectives, process or local operations, and b) many factors found in the literature are insufficiently defined for immediate use. We extended our model to include nodes with the new kinds of contextual meaning.

### 4.1 Categorising Literature Studies

The next step in the evolution of our model was to extend the literature categorisation with the aim of more extensively testing the expanded model. We sourced titles and abstracts from each of:

- Elsevier's Scopus for IS technical and social sciences literature
- Academic Search Premier (EBSCO) for business focused material

Our search spanned 2014-2015. The numbers for candidate documents are shown in table 2. The original plan was to process all candidate documents. We took a 'coverage' approach, in that we early on selected sources that appeared to vary in topic and date. However, once we had processed 30 or so documents from Scopus, it seemed that no new kinds of factor were appearing. We moved to EBSCO documents in case these provided some variation, but this was not the case. According to Routio, the indication that a study has reached a point of data saturation is that the "study no longer reveals new interesting information" (Routio, 2007). We felt that the gains in completing all documents would probably not be significant, and made a pragmatic decision to alter our approach. Our justification was that we were using the literature to *test* a model (as opposed to creating a model from the literature) and the research was exploratory i.e. more formal refinement would follow. We processed 50 documents, as described below, from the Scopus and EBSCO sources. This resulted in a modification to the base dimensions of our model. We then tested the resulting modified model by processing a further 12 documents from Google Scholar, applying the same search string.

For each source, a second pass involved reading the text of a candidate document in sufficient depth to ascertain whether the document was relevant. This was carried out by the first author, with the second author performing 'random' checks on decisions made. The numbers of documents included in the study are shown in the *Included* rows in table 2.

From each of the included documents, we extracted into a dedicated document words or terms that could be viewed as stating or describing a contextual factor. As in the pilot, our strategy was to be as comprehensive as possible in our identification of contextual factors. This meant that we wanted to expose factors that may not be typically considered as context. For example, in Section 1, we noted that the software-as-a-service paradigm has revealed the need for different kinds of practice, but this is not normally viewed as a contextual factor. We thus chose to include studies that contain any thoughts or description about what might affect practice efficacy. We did not evaluate the studies in which the elements were mentioned for quality. We also did not 'tidy up' the found elements by making value judgements about whether two elements had the same meaning. We felt that such evaluations would effectively remove some of the nuances of identification and would thus compromise our efforts. The underlying issue here is one of a lack of common, agreed vocabulary for software projects.

We then accumulated terms into a spreadsheet, with one page for each of the base model dimensions, and one for each of the 'other' classifications to be examined i.e. 'Secondary', 'Ambiguous', 'Strategy', 'Objectives' and 'Process'. During this stage, an attempt was made to remain true to the meaning of terms used. However, some interpretation was unavoidable as we attempted to categorise terms *within* dimensions. We applied thematic analysis to categorise *within* dimensions.

The outcome of this step was the discovery of factors that initiated a change to the base structure to include the four dimensions (see figure 1):

**People.** Cultural characteristics affecting peoples' ability to perform

**Place.** Peoples' availability affecting logistics and communications

**Product.** Characteristics of the product that is being developed

**Process.** Processes external to the initiative (as com- pared with practices within the initiative).

An example is the term 'legal'. To illustrate, a practice which appears to have good fit with the model may be disallowed because it requires sharing of intellectual property and there is no appropriate agreement in place. This example represents a restriction resulting from processes external to the software development group. On further thought, we realised that such restrictions might come also from, for example, process-related expectations of the parent organisation. We required a general idea of constraints on practice that result from processes external to the initiative. We extended the *how* dimension to refer to any constraints on practice implementation resulting from processes external to the initiative.

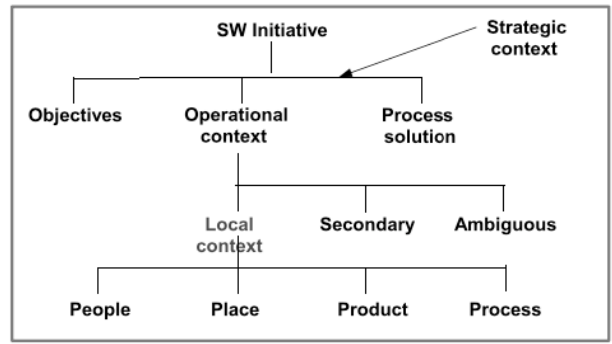

Figure 1: Software context elements.

Table 3: Local operational context factors.

| | | |
|---|---|---|
| People | Entity | Capability |
| | | Motivation |
| | | Empowerment |
| | | Cultural cohesion |
| | Interface | Cultural cohesion |
| Place | Entity | Physical distance |
| | | Temporal distance |
| | | Availability |
| | Interface | Physical distance |
| | | Temporal distance |
| | | Availability |
| Product | Product type | e.g. embedded |
| | Lifecycle stage | e.g. new, mature |
| | Standards | e.g. safety |
| | Requirements | e.g. emergent, complete |
| | Implementations | e.g. modularity |
| Process | Client | |
| | Parent org | |
| | Legal | |
| | Financial | |

In the earlier model, one of the categories within the *how* dimension was 'Client Demographic'. The original viewpoint was that demographic would affect specification and delivery mechanisms i.e. 'how' the product was defined and delivered. We realised the demographic might also affect logistics (for example, 'global market') and culture (for example, 'government agency') and we reclassified as *Secondary*. During this re-evaluation, we understood that the *when* dimension i.e. relating to the lifecycle stage of the situated product, is a product-related constraint and as such can be merged with *what*. The model was now as shown in figure 1 and table 3.

### 4.2 Industry Evaluation

In this section, we overview our efforts to represent local operational context for two small industry initiatives. We wanted to determine if the model was usable in practice. Our approach was to interview senior members of these organisations, applying a questionnaire based on our context-model-under-test.

Table 4: Org A - Contextual values for Product.

| Type | Migration |
|---|---|
| Lifecycle stage | Mature |
| Standards | None applicable |
| Specification | Well-understood |
| As Implemented | Modular code base |

Table 5: Org A - Contextual values for People.

| PM-capability | High |
|---|---|
| PM-motivation | High |
| PM-empowerment | Is owner and fully empowered |
| Team-capability | Mixed |
| Team-motivation | Low (background project) |
| Team-empowermt | High (own decisions) |
| Team-cohesion | High (common understanding) |
| Client-capability | Low (no longer available) |
| Client-motivation | N/A - unaware of migration |
| Client-empowermt | N/A - unaware of migration |
| Team-client-cohes | N/A - unaware of migration. |

Table 6: Org B - Contextual values for Product.

| Type | Stand-alone |
|---|---|
| Lifecycle stage | New Product Development (NPD) |
| Standards | NZ standards for roofing design |
| Specification | Uncertain as new innovation |
| As Implemented | N/A as this is a new product |

Table 7: Org B - Contextual values for People.

| Team-capability | All experienced in technologies Low application area knowledge |
|---|---|
| Team-empowermt | High |
| Team-motivation | High (well-paid team) |
| Team-sharing | High (same room) |
| Client-capability | Expert in application area |
| Client-empowermt | High (owns company) |
| Client-motivation | High (wants purpose built app) |
| Team-client-cohes | Low (no mutual understanding) |

### 4.2.1 Organisation A

Organisation A has been producing custom EDI (Electronic Document Interchange) software for a global marketplace since 1984. Client systems are customised, based on a core, and are manned by administrators. This means that technical representatives are no longer available at the client end. The project studied was an internal project to migrate one version of a core system to a new upgraded version. The aims were a) to increase productivity for clients by improving efficiency, b) to effect risk mitigation by moving from an outdated technology to one that is more scaleable and less limited, and c) to take the opportunity to establish standards to support future projects. Clients were largely unaware the migration was taking place. Midway through the project, it be- came clear that progress was slow due to a lack of urgency and *completion time* was included as an objective.

Project members included a manager, owner/analyst, lead developer and 2-4 further developers. The manager was highly motivated to empower the team to 'own' decisions and to 'future proof' the product against future inexperienced developers. The developers varied in experience levels and lacked application area knowledge. We interviewed the manager and the owner/analyst in a single interview of 45 minutes. Context values for Product and People are shown in tables 4 and 5. For the Place dimension, project members are located on the same floor in the same building — effectively in the same room. Sometimes developers worked from home, but were available. There were no external constraints on Process — the senior team member was the owner and, as an internal process, client processes had no bearing.

### 4.2.2 Organisation B

Organisation B is owned and managed by an experienced civil engineer, who recognised a need for software for roof design, for example, to support optimisation of materials required and to produce detailed invoices. A project to explore this idea was set up. As the manager was unfamiliar with roof design, a decision was made to work with an experienced roofer and to deliver an initial version of the system to him. However, the intention to later expand to other kinds of roof and to an international market was present throughout. and the aim was to spend time in the short term to achieve productivity savings later. The team comprised the owner/manager and two highly experienced, contracted developers who worked in the same room. The client was available for specification and feedback. We interviewed the owner-manager in a 50 minute session.

The project was strongly driven by the owner/manager who had responsibility for long-term planning, short-term scoping and delivering to the roofing expert. His vision of 'implementing within the bigger picture' and experiences as an engineer also impacted on how implementation was carried out. The interview tended to be free-flowing with prompts to ensure focus.

The main objectives identified for the project were client satisfaction and extendible code base. Contexts are presented in tables 6 and 7. The team was small and worked full-time from the project office. No external constraints from external processes were identified.

### 4.2.3 Industry Practices

The interviewees in both organisations were asked about the practices they felt had been most helpful or unhelpful in meeting objectives. We will present and discuss those findings in a later work. However, to illustrate how a picture of a situated practice might be built up, we show in table 8 the results for the practice 'create coding standards'. The practice is indicated for objectives 'Quality' and 'Create extendible code base', if the people involved are capable and not if the initiative is time constrained. We have the beginnings of an evidence base for the practice of creating coding standards.

Table 8: Practice - create coding standards.

| Indicated | Objectives | quality (A) |
|---|---|---|
|  |  | extendible code base (B) |
|  | People: | capable (B) |
| Contra | Process: | time constraints (A) |

## 5. DISCUSSION AND LIMITATIONS

One of the main findings during this research was that many terms are used without a clear definition of what they *mean*. For example, during analysis of one of the test documents (Wallace and Keil, 2004), we observed that the risk framework applied resulted in identification of several 'risks' which we would classify as *ambiguous*. If a risk in not clearly articulated, it seems clear that any mitigation attempts are likely to be less than effective.

An assumption of the model is that the characteristics of individuals are subsumed by team characteristics and this needs to be tested. For example, can we characterise a team as of 'Medium capability' when a) some members are capable and others are of low capability, and b) all are of medium capability. This is an open research question. We have also assumed that team size on its own is not relevant, but is characterised by 'shared understanding' and 'capability' (a well-run large team may be more capable that a dis- functional small team). This is another assumption to be tested.

The primary limitation for this study is the risk of subjectivity. For reasons of limited resourcing, most of the work has been carried out by the primary author, with ad-hoc checks by the second author. Aspects of this risk include the following:

- There is clearly a tendency to view each identified factor from the perspective of the proposed model and this may mean that other, possibly more useful, perspectives are missed. For example, a study about 'Adaptive software' — was accepted as relevant only because we had a *product* dimension in the model and recognised the product type as relating to that dimension. We do not see any way to avoid this, as the selection of dimensions *is* the model we are exploring.

- Meanings may have been lost or mis-interpreted as terms were transported from the dedicated summary to the analysis spreadsheet. In mitigation, we have all data available.

Both teams in the industry trial were small and this is clearly a serious limitation.

## 6. SUMMARY AND FUTURE WORK

In this paper, we have described the evolution of a model for software process context to support evidence accumulation for situated software practices.

The study represents the first (exploratory) stage in a three stage process to conceptualise, refine and apply a model for context (Routio, 2007). We justified our approach of starting from a conceptual structure rather than from the existing literature by observing that a) existing literature-based models tend to be unclear as to category *properties*, b) the possibly infinite number of contextual factors means that we can never be certain that all have been found, and c) we do not know what paradigms will appear in the future. Model evolution involved testing by categorising studies from the literature and a small industry trial involving two organisations, where we captured key contexts for two projects.

Our research philosophy is one of *pragmatism* (Creswell, 2014). The pragmatist considers theories as "the products of a consensual process ... to be judged for their utility" (Easterbrook et al., 2008). We have created "a starting point (e.g. a framework) that identifies aspects of a topic" (Stol and Fitzgerald, 2015).

Key findings thus far are that contextual factors as named in the literature have different kinds of *meaning* and many are too vague to be of use for process tailoring. We have identified the kinds of meaning as relating to organisational strategy, project objectives, practice implementation and local operational context. Local context dimensions are *People*, *Place*, *Product* and *Process*. We explored the internal structures of these dimensions, understanding that, if overly complex, the proposed model will be unworkable in practice.

We accept that the research quality of this paper is reduced as a result of the risk of subjectivity and the limited nature of the study.

Our contributions are a deeper understanding of the different kinds of *meaning* represented by 'context' factors as stated in the literature and a model that is sufficiently complete for use as a basis for future refinement. We will next formally refine the model in conjunction with others. We are currently developing a taxonomy based on the model to support discussion with researchers and practitioners.

## REFERENCES

Avison, D. and Pries-Heje, J. (2008). Flexible information systems development: Designing an appropriate methodology for different situations. In *Enterprise information systems : ICEIS 2007*, pages 212–224. Springer.

Basili, V. R., Shull, F., and Lanubile, F. (1999). Building Knowledge through Families of Experiments. *IEEE Trans. on SW Engineering*, 25(4):456–473.

Bjarnason, E., Smolander, K., Engstrom, E., and Runeson, P. (2015). A theory of distances in software engineering. *Inf. and SW Technology*, 70(C):204–219.

Bosu, M. F. and MacDonell, S. G. (2013). A Taxonomy of Data Quality Challenges in Empirical Software Engineering. In *Proc. ASWEC 2013*, pages 97–106.

Campanelli, A. S. and Parreiras, F. S. (2015). Agile methods tailoring - A systematic literature review. *Journal of Systems and Software*, 110:85–100.

Carver, J., Voorhis, J. V., and Basili, V. (2004). Understanding the Impact of Assumptions on Experimental Validity. In *Proc. ISESE'04*, pages 251–260. IEEE.

Clarke, P., Mesquida, A.-L., Ekert, D., Ekstrom, J., Gornostaja, T., Jovanovic, M., Johansen, J., Mas, A., Messnarz, R., Villar, B. N., O'Connor, A., O'Connor, R. V., Reiner, M., Sauberer, G., Schmitz, K.-D., and Yilmaz, M. (2016). An Investigation of Software Development Process Terminology. In *SPICE 2016)*, volume 609 of *CCIS*, pages 351–361. Springer.

Clarke, P. and O'Connor, R. V. (2012). The situational factors that affect the software development process: Towards a comprehensive reference framework. *Inf. and SW Technology*, 54:433–447.

Creswell, J. W. (2014). *The Selection of a Research Approach*, pages 31–55. Sage Publications Inc.

Dingsøyr, T. and Lassenius, C. (2016). Emerging themes in agile software development: Introduction to the special section on continuous value delivery. *Information and Software Technology*, 77:56–60.

Dybå, T., Sjøberg, D. I., and Cruzes, D. S. (2012). What Works for Whom, Where, When and Why? On the Role of Context in Empirical Software Engineering. In *Proc. ESEM 2012*, pages 19–28.

Easterbrook, S., Singer, J., Storey, M., and Damian, D. (2008). Selecting empirical methods for software engineering research. In *Guide to Advanced Empirical Software Engineering*, pages 285–311. Springer.

Jacobson, I., Meyer, B., and Soley, R. (2013). Software Engineering

Method and Theory. http://www.semat.org. Kirk, D. and MacDonell, S. G. (2014a). Categorising soft- ware contexts. In *Proc AMCIS 2014*.

Kirk, D. and MacDonell, S. G. (2014b). Investigating a conceptual construct for software context. In *Proc. EASE 2014*, number 27.

Kirk, D. and MacDonell, S. G. (2016). An Ontological Analysis of a Proposed Theory for Software Development. In *Software Technologies - ICSOFT 2015*, volume 586 of *CCIS*, pages 1–17. Springer.

Lengnick-Hall, C. A. and Griffith, R. J. (2011). Evidence-based versus tinkerable knowledge as strategic assets: A new perspective on the interplay between innovation and application. *Journal of Engineering and Technology Management*, 28:147–167.

MacCormack, A., Crandall, W., Henderson, P., and Toft, P. (2012). Do you need a new product- development strategy? *Research Technology Management*, 55(1):34–43.

Munezero, M., Yaman, S., Fagerholm, F., Kettunen, P., Mäenpää, H., Mäkinen, S., Tiihonene, J., Riungu-Kalliosaari, L., Tuovinen, A.-P., Oivo, M., Münch, J., and Männistö, T. (2017). *Continuous Experimentation Cookbook*. DIMECC Oy, Helsinki, Finland.

Orlikowski, W. (2002). Knowing in Practice: Enabling a Collective Capability in Distributed Organizing. *Organization Science*, 13(3):249–273.

Petersen, K. and Wohlin, C. (2009a). A comparison of is- sues and advantages in agile and incremental development between state of the art and an industrial case. *Journal of Systems and Software*, 82:1479–1490.

Petersen, K. and Wohlin, C. (2009b). Context in Industrial Software Engineering Research. In *Proc. ESEM 2009*, pages 401–404. IEEE.

Routio, P. (2007). Models in the Research Process. http://www2.uiah.fi/projects/metodi/177.htm.

Runeson, P., Stefic, A., and Andrews, A. (2014). Variation factors in the design and analysis of replicated con- trolled experiments. *Empirical Software Engineering*, 19:1781–1808.

Sjøberg, D. I., Dybå, T., Anda, B. C., and Hannay, J. E. (2008). Building Theories in Software Engineering. In et al., F. S., editor, *Guide to Advanced Empirical Software Engineering*, pages 312–336. Springer-Verlag.

Sjøberg, D. I., Hannay, J. E., Hansen, O., Kampenes, V. B., Karahasanovic, A., Liborg, N.-K., and Rekdal, A. C. (2005). A Survey of Controlled Experiments in Software Engineering. *IEEE Transactions on Software Engineering*, 31(9):733–753.

Stol, K.-J. and Fitzgerald, B. (2015). Theory-oriented software engineering. *Science of Computer Programming*, 101:79–98.

Stuckenberg, S. and Heinzl, A. (2010). The Impact of the Software-as-a-Service concept on the Underlying Software and Service Development Processes. In *Proc PACIS 2010*, pages 1297–1308.

Turner, R., Ledwith, A., and Kelly, J. (2010). Project management in small to medium-sized enterprises: Matching processes to the nature of the firm. *International Journal of Project Management*, 28:744–755.

Wallace, L. and Keil, M. (2004). Software Project Risks and their Effects on Outcomes. *Comm. ACM*, 47(4):68–73.

Wang, X., Conboy, K., and Cawley, O. (2012). Leagile software development: An experience report analysis of the application of lean approaches in agile soft- ware development. *Journal of Systems and Software*, 85:1287–1299.

Zachman, J. A. (2009). Engineering the Enterprise: The Zachman Framework for Enterprise Architecture. http://www.zachmaninternational.com/index.php/the-zachman-framework.